\documentclass[%
 reprint,
superscriptaddress,
 amsmath,amssymb,
 aps,
]{revtex4-2}

\usepackage[justification=justified]{subcaption}
\usepackage[colorlinks=true, linkcolor=blue, citecolor=black, urlcolor=blue]{hyperref}
\usepackage{bm, booktabs, dcolumn, graphicx, ragged2e, rotating}

\DeclareCaptionJustification{reallyjustified}{\justifying}
\begin{document}

%TC:ignore

\preprint{APS/123-QED}

\title{New Bounds on Exotic Long-Range Spin-Spin Interactions}

\author{N.B. Clayburn}
\affiliation{%
 Department of Physics \& Astronomy, Amherst College, Amherst, Massachusetts 01002, USA 
}%
\affiliation{Department of Physics, Keene State College, Keene, New Hampshire 03435, USA}
\author{A. Glassford}
\author{T. Uelmen}
\author{A.R. Kyung}
\author{Y. Boneva}
\author{S. Salim}
\author{A.S. Weiss}
\author{F. Waldherr}
\author{C. Carlin}
\author{S.K. Peck}
\author{L.R. Hunter}
\email{lrhunter@amherst.edu}
\affiliation{%
 Department of Physics \& Astronomy, Amherst College, Amherst, Massachusetts 01002, USA 
}%

\date{\today}

\begin{abstract}
Many proposed extensions to the Standard Model of particle physics introduce new bosons that can mediate forces which couple to particle spin. Here we describe a search for such forces coupling spin-polarized neutrons and protons in our magnetometer to spin-polarized electrons within Earth. We measure these interactions by varying the orientation of an optical $^{199}$Hg-$^{133}$Cs free-precession comagnetometer mounted upon a precision rotation platform. From these measurements, we establish upper bounds on the dimensionless coupling constants associated with the axial-axial potential $V_2$ and the axial-vector potential $V_{11}$ as a function of the force's range $\lambda$. For the electron-neutron and electron-proton potential $V_2$ at infinite range, we find $|g_A^eg_A^n| \leq 3.0 \times 10^{-48}$ and $|g_A^eg_A^p| \leq 3.0 \times 10^{-47}$. For $V_{11}$, we find our most stringent bounds to be $|g_A^eg_V^n| \leq 2.2 \times 10^{-25}$ and $|g_A^eg_V^p| \leq 2.2 \times 10^{-24}$ at $\lambda \approx 10^3$ km. Our results represent an improvement over previous results by up to a factor of 17 and set the most stringent bounds on long-range axial-axial and axial-vector couplings between electron spins and neutron and proton spins.
\end{abstract}

\maketitle

%TC:endignore

\section{Introduction}

Many proposed extensions to the Standard Model predict the existence of new bosons which mediate interactions between fermion spins. Axions \cite{axionOriginalPaper} and axion-like particles \cite{alpsPaper} are examples of proposed spin-0 bosons, while $Z'$ bosons, dark photons, and paraphotons \cite{darkPhotonTheory,paraphotonTheory} are examples of proposed spin-1 bosons. All of these particles are potential dark-matter candidates. In addition, various torsion gravity models also predict long-range interactions between spins \cite{kostolecky1998,hammond2002,flambaum2009}.

The assumptions of rotational invariance, gauge invariance, and conservation of mass-energy lead to a classification of 16 possible spin-dependent interaction potentials \cite{dobrescu2006, safronova2018}. Here we adopt the conventions of Ref. \cite{budker2025} and restrict our discussion to the dot product and cross product potentials \begin{eqnarray}
    V_2 &=& -g_A^Xg_A^Y\frac{\hbar c }{4 \pi} (\bm{\sigma}_X \cdot \bm{\sigma}'_Y )\frac{1}{r}  e^{-r/\lambda} \\
    V_{11}|_{AV} &=& -g_A^X g_V^Y \frac{\hbar^2}{8\pi m_Y} \nonumber \\ & &    (\bm{\sigma}_X \times \bm{\sigma}'_Y) \cdot \hat{r}\left(\frac{1}{r^2}+\frac{1}{\lambda r} \right) e^{-r/\lambda}, 
\end{eqnarray} where $g$ denotes the vector ($g_V$) or axial ($g_A$) coupling constants of fermions $X$ or $Y$ with mass $m$ and spin $\bm{\sigma}$. The interaction range of the force is given by $\lambda=\frac{\hbar}{m_Z c}$, where $m_Z$ is the mass of the mediating boson.

Ref. \cite{budker2025} provides a thorough review of the field as of early 2025. We restrict our focus here to long-range ($\lambda \geq 1$ km) electron-neutron ($e$-$n$) and electron-proton ($e$-$p$) interactions. In 2013, using Earth's mantle and crust as a source of spin-polarized electrons allowed for stringent new bounds on long-range spin-spin interactions (LRSSIs) \cite{hunter2013,ang2014}. Recently, multiple terrestrial experiments were re-interpreted in the context of a spinless Earth model \cite{clayburn2023} and using the moon and sun as sources \cite{wu2023}. An effort to extract bounds on exotic forces from the Earth-spin model and measurement of the precession of the perihelion of Earth \cite{poddar2023} was shown to be of limited sensitivity \cite{noNetSpin}. Searches for exotic spin-spin and spin-mass couplings using space-borne comagnetometers and Earth as a spin and mass source have recently been proposed \cite{huang2025,lai2026}.

\section{Methods}

\subsection{Overview}

Here we describe a new experiment that uses the spin-polarized electrons within Earth as a source and a free-precession $^{199}$Hg-$^{133}$Cs comagnetometer as a detector \cite{kimball2017,romalisHg}. This free precession geometry greatly suppresses the vector AC light effects that limited the sensitivity of the earlier Hg-Cs experiment \cite{hunter2013}.

In our comagnetometer, we compare the responses of three stacked magnetometer cells. Two $^{133}$Cs cells flank a single $^{199}$Hg cell, as shown in Figure \ref{fig:apparatus}. The Cs magnetometers are primarily sensitive to effects associated with their electron spin. Because of the phenomenal bounds on electron-electron LRSSI effects that were extracted \cite{hunter2013} from the Seattle torsion-pendulum experiment \cite{seattleExperiment}, we do not anticipate any significant Cs response to the relevant exotic interactions at our level of precision. Hence, any anomalous LRSSI response in our apparatus can be attributed to the $^{199}$Hg nucleus and will be used for placing bounds on the interactions between Earth's unpaired, polarized electron spins and $^{199}$Hg nucleon spins. Subtracting the average Cs magnetic field measurement from the Hg ``effective" magnetic field measurement, we create the ``Hg-Cs difference". This signal is relatively insensitive to changes in the magnetic field. However, if exotic spin-dependent interactions between Hg atoms and Earth exist, they will slightly modify this difference signal. Furthermore, the difference signal associated with an interaction between Hg's spins and Earth must change sign when the direction of the magnetic field is reversed. In the experiment, we monitor the Hg-Cs difference field as a function of the orientation of the applied magnetic field with respect to Earth. The different magnetic field orientations are realized by rotating the entire apparatus on a precision rotation table to precisely align the applied magnetic field with the fixed cardinal directions.

\subsection{Magnetometer}

Most of our data is collected with an applied field of 1 $\mu$T. Because the Cs precession frequency ($\sim$3.5 kHz) is much higher than that of Hg ($\sim$7 Hz), the mechanics of optical pumping differ between the two atomic species. For Cs, we use an acousto-optical modulator (AOM) to modulate the intensity of a circularly-polarized pump beam (894 nm) at the Cs Larmor frequency with a 20\% duty cycle for 0.25 seconds. The pump laser frequency is centered 80 MHz above the $F=3 \to F'=4$ hyperfine transition. When the atomic polarization nears its maximum, we extinguish the pump and interrogate the precession with a linearly-polarized probe laser tuned approximately 2.5 GHz below the $F=4 \to F'=3$ resonance. To suppress tensor light shifts, the angle between the probe's linear polarization and the magnetic field is set to the ``magic" angle \cite{peck2016}.

For Hg, we use a commercial UV frequency-quadrupled CW laser for both pumping and probing. We split our 254 nm beam into a bright circularly-polarized pump beam and a weak linearly-polarized probe beam. Each beam passes through a mechanical shutter. For the pump cycle, we tune the laser to the $6^1S_0 \to 6^3P_1$ transition. The probe shutter is closed, and the pump shutter modulates at the Hg Larmor frequency with a 20\% duty cycle. After 40 seconds, we attain a significant polarization and close both shutters for $\sim2$ seconds as we tune the laser approximately 8 GHz above the resonance. We then open the probe shutter.

Our $^{133}$Cs pump laser is locked to a Cs saturated absorption cell, guaranteeing optical frequency stability to within 30 kHz. A portion of that light is used to stabilize a Fabry-Pérot cavity. All other optical frequencies in the experiment are fixed by locking to the frequency-stabilized Fabry-Pérot transfer cavity. Both the Hg laser and Cs pump laser are intensity-stabilized.

For both Cs and Hg, the free-precession of the spins induces optical rotation of the linear polarization of the weak, far off-resonance probe beams. Optical polarimeters monitor this rotation, which appears as an exponentially decaying sine wave. We fit this function to determine the precession frequency in each of the three cells.

The spin-relaxation times, $T_2$, of Hg and Cs differ dramatically. For Cs, $T_2 \sim 90$ ms, while for Hg, $T_2 \sim 20$ seconds. Due to the difference in $T_2$, we measure 120 Cs pump-probe cycles (each taking 0.5 seconds) for each Hg probe cycle (60 seconds).

\subsection{Apparatus}

\begin{figure}[ht]
    \centering
    \includegraphics[width=1\linewidth]{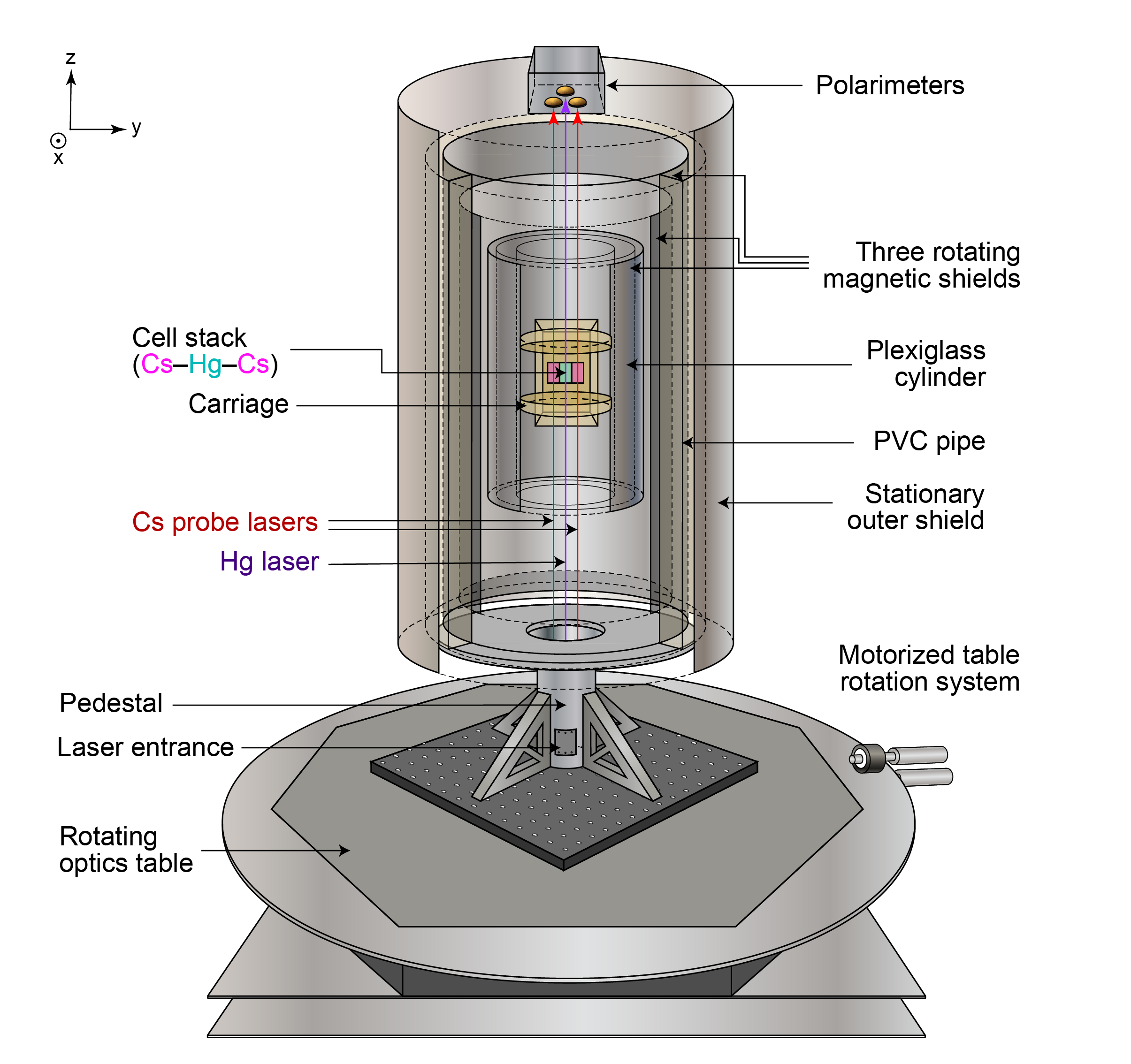}
    \caption{Schematic of experimental apparatus (not to scale). The pump laser beams (not pictured) propagate nearly parallel to the probes such that they overlap near the center of the cells. To avoid saturating the polarimeters, the pump beam is tilted slightly along the $x$-axis. Optical elements are not shown.}
    \label{fig:apparatus}
\end{figure}

Figure \ref{fig:apparatus} shows a schematic of our apparatus. The heart of our apparatus is the trio of atomic vapor cells housed within a set of magnetic field coils. This assembly is mounted on a rotatable mechanical superstructure.

The atomic vapors are contained in rectangular cells, fabricated by attaching two optical flats to the ends of a $1"$ length of $\frac{1}{2}"\times 1"$ rectangular tubing. The flats allow for undistorted passage of the laser beams. The isotopically-enriched $^{199}$Hg cell is composed of quartz with an inner coating of dotriacontane, a wax which reduces the rate of depolarization due to wall collisions. 400 Torr of CO is added to the Hg cell to minimize wall collisions and quench the $^{199}$Hg $6^3P_0$ metastable state \cite{romalisHg}. The $^{133}$Cs cells are fabricated from borosilicate glass. An internal coating of alkene 1-nonadecene reduces spin relaxation on the walls \cite{waxCoating}. All three cells are operated at room temperature.

The trio of vapor cells is mounted in a plexiglass carriage. That carriage is mounted in an acrylic cylinder embedded with magnetic field coils. Three internal coils provide homogeneous fields along $\hat{x}$, $\hat{y}$, and $\hat{z}$. Our main precession field points along the $x$-axis. Two other coils are quadrupolar field coils, which allow for control of the linear gradients along the $\hat{x}$ and $\hat{y}$ directions. The acrylic cylinder is mounted inside a three-layer $\mu$-metal magnetic shield assembly, which is mounted within a large PVC pipe. This assembly sits on a pedestal mounted on an octagonal optics table. This table can be rotated with a reproducibility of 0.004$^\circ$. A fourth, stationary $\mu$-metal shield, suspended from the ceiling, surrounds the inner assembly and shields it from Earth's magnetic field.

Flanking the entire apparatus is a pair of square coils, 9 feet on each side, wrapped in a Helmholtz-like configuration. These coils produce a reasonably homogeneous magnetic field in the region of the apparatus. They are designed to cancel the North component of Earth's geomagnetic field during data acquisition. They also provide a means to measure the attenuation ratio of our 4-shield assembly, which we find to be $\sim 3 \times 10^{6}$.

\subsection{Procedures}

\emph{Data Collection.} A data point consists of a single Hg frequency measurement over one pump-probe cycle of $^{199}$Hg and concurrent Cs frequency measurements taken over 120 pump-probe cycles of $^{133}$Cs. We take the inverse-variance weighted average across the 120 Cs measurements, then convert all three atomic precession frequencies into magnetic fields by dividing by their gyromagnetic ratios. To approximate the field at the Hg cell, we average the magnetic field measured by the two Cs cells. We subtract this averaged Cs magnetic field from that measured in the Hg cell to form a single data point. We take five data points in a given cardinal direction before rotating to the opposite direction and repeating this process. For North (N) and South (S), the order in which we rotate to different positions is N-S-S-N-S-N-N-S (and similarly for East and West). We call one of these sets an ``octet." Each octet represents approximately 2 hours of data collection. We take the weighted average measurement across the octet's North data and the weighted average measurement across the octet's South data. We then subtract South from North to recover an octet-level measurement. Our time ordering is chosen so that, when averaging across an octet, first- and second-order drifts in time are eliminated. We alternate between taking North-South and East-West octets. We then perform a weighted average across all North-South octets and all East-West octets to form our final result. We calculate the uncertainty in our final values using the weighted variance of the mean \cite{weightedavg}. We express all of our results in terms of the equivalent Hg frequency $f_{\text{Hg}}=\gamma_{\text{Hg}} B$.

\emph{Systematic Errors.} For most of our systematic errors, prior to the data collection process, we form a slope by intentionally varying each parameter of our experiment and recording the resulting impact on our science signal. During the data collection process, we measure the actual variation in each parameter upon table reversal. We then multiply the slope by the actual variation to compute an estimated impact on the science signal. Finally, we subtract the estimated impact away, propagating uncertainties in quadrature. We correct for 13 systematic errors on an octet-by-octet basis. We report the effect of these ``local" systematics in Appendix \ref{app:localSystematics}. 

We also investigated five ``global" systematic errors on an experiment-wide basis. These global errors could theoretically have an impact on the entire dataset. We report these systematics in Appendix \ref{app:globalSystematics}.

\emph{Blind.} All data was collected under a blind to reduce experimenter bias. An unknown random value between +760 Hg nHz and -760 Hg nHz, the bounds placed by the last iteration of this experiment \cite{hunter2013}, was applied to each octet. All data cuts and selection were done before removing the blind.

\section{Results}

\begin{figure*}[ht!]
    \centering
    \includegraphics[width=1\linewidth]{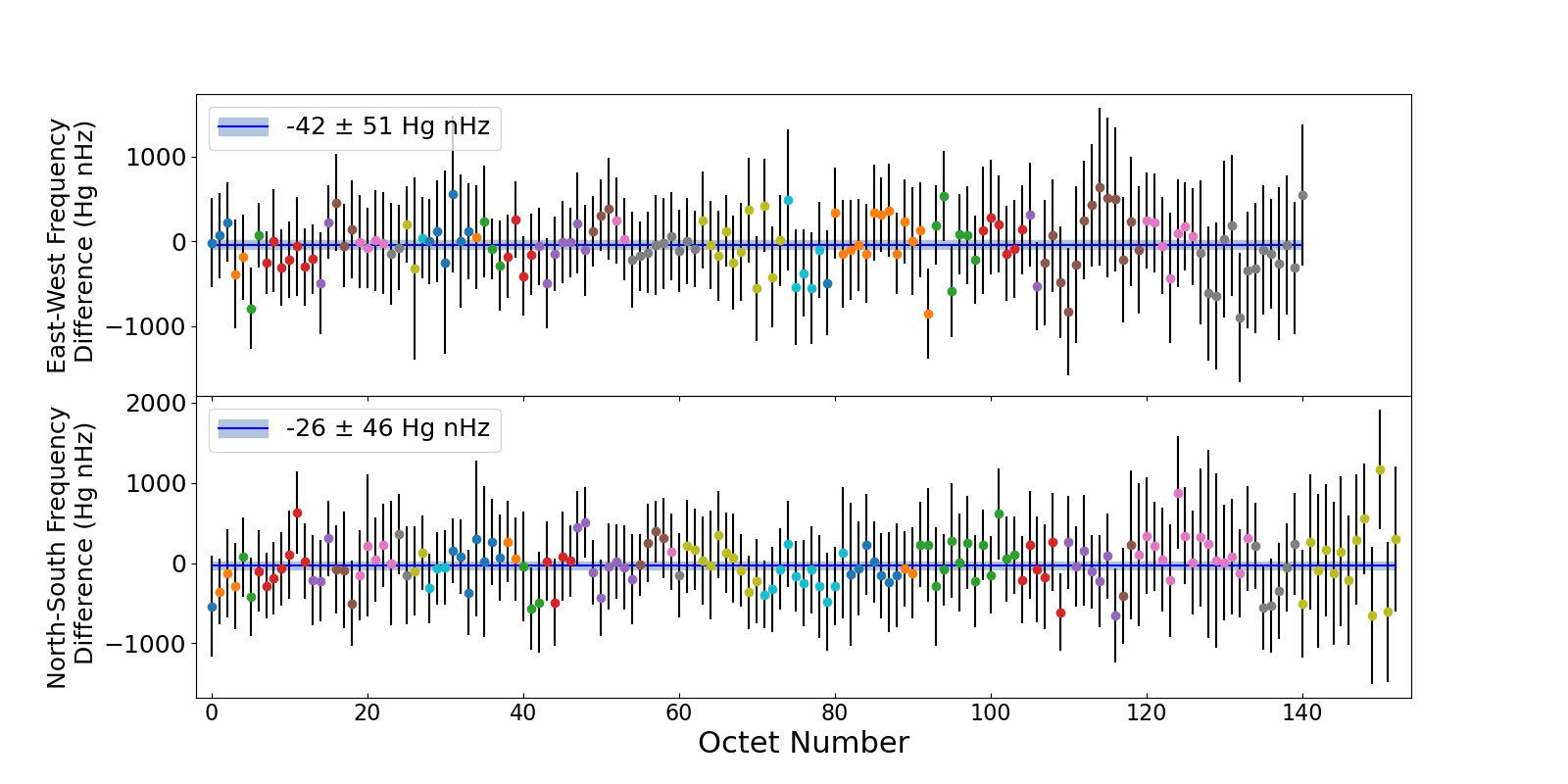}    
    \caption{Unblinded final results organized by octet. Local and global systematic corrections have been incorporated. Each data point represents an octet with 2$\sigma$ error bars. The blue horizontal line and shaded region represent the averages and confidence intervals respectively. Runs consisting of data collected sequentially are differentiated by color. Some runs only contain data from either the North-South or East-West directions.}
    \label{fig:octetFinalAnswers}
\end{figure*}

After all collaborators reviewed the data and analysis, unanimous consent was given to unblind the experiment. Figure \ref{fig:octetFinalAnswers} shows our unblinded result. All uncertainties are given as 2$\sigma$. With all known systematics taken into account, we measure a N-S difference of $-26 \pm 46$ Hg nHz, and an E-W difference measurement of $-42\pm 51$ Hg nHz. As these are both consistent with null results, we interpret them as bounds on exotic forces. We derive an upper bound on North-South of $72$ Hg nHz and an upper bound on East-West of $93$ Hg nHz. We also analyze our data in different configurations in Appendix \ref{app:crossChecks}. We find no significant difference between various configurations.

We convert to energy bounds by multiplying by $h$ and dividing by 4 to account for two sets of doubling: once from the subtraction of North-South (and East-West) and once from the comparison of two equal and opposite spin states (+1/2 vs -1/2). We use the formula
\begin{equation}
    \beta_{N/E}=\frac{h\Delta f_{N/E}}{4}
\end{equation} to derive orientation-dependent energy shifts $\beta$, with subscripts denoting orientation along North-South (N) or East-West (E). Using the nuclear spin contribution calculations from Ref. \cite{hgNuclearTheory}, we divide by 0.302 for neutrons ($n$) and 0.031 for protons ($p$) to derive energy bounds $\beta_{N/E}^{p/n}$. Table \ref{tab:unblindedResults} shows our results. Compared with the earlier experiment \cite{hunter2013}, the North-South bounds represent an improvement of a factor of approximately 17, while the East-West bounds improve by a factor of approximately 9.

\begin{table}[ht]
    \captionsetup{justification=reallyjustified, singlelinecheck=true}
    \centering
    \resizebox{\columnwidth}{!}{
    \renewcommand{\arraystretch}{1.2}
    \begin{tabular}{|l|c|c|} \hline
        Signal & North-South & East-West \\ \hline
        Raw Data & $-15 \pm 45$ Hg nHz & $-44 \pm 47$ Hg nHz \\
        Local Corrections & $-26 \pm 46$ Hg nHz & $-43 \pm 47$ Hg nHz \\
        Global Corrections & $-26\pm 46$ Hg nHz & $-42 \pm 51$ Hg nHz \\ \hline
        Frequency Bounds & 72 Hg nHz & 93 Hg nHz \\
        Energy Bounds & $\beta_N=7.49\times 10^{-23}$ eV & $\beta_E=9.59 \times 10^{-23}$ eV \\ \hline
        Neutron Bounds & $\beta_N^n=2.48
        \times 10^{-22}$ eV & $\beta_E^n=3.18 \times 10^{-22}$ eV \\
        Proton Bounds & $\beta_N^p=2.42 \times 10^{-21}$ eV & $\beta_E^p=3.09 \times 10^{-21}$ eV \\ \hline
    \end{tabular}
    }
    \caption{Unblinded Results. The first row shows our result without any systematic corrections applied. The second row has local systematic corrections applied. The third row has both local and global systematic corrections applied. The frequency bound is the absolute value of the measured frequency plus uncertainty.}
    \label{tab:unblindedResults}
\end{table}

\begin{figure*}[t]
    \centering
    \includegraphics[width=1\linewidth]{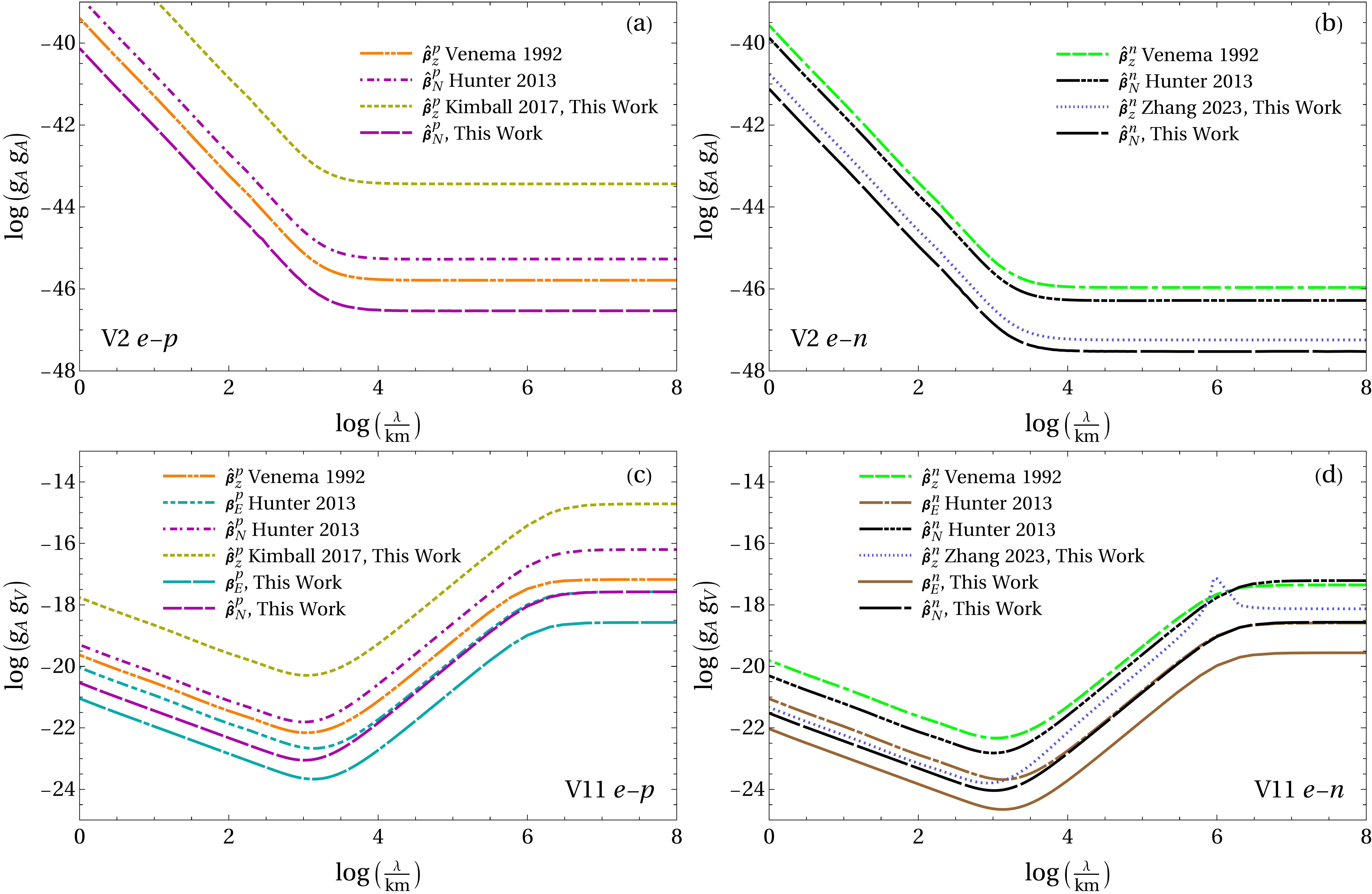}
    \caption{Bounds on the dimensionless coupling constants associated with the $V_2$ and $V_{11}$ potentials, plotted against the interaction range $\lambda$. These bounds are extracted from the $\beta$ indicated by coloring and dashing. We follow the notation of Ref. \cite{heckel2008} in using $\hat{\beta}$ to indicate that a correction for Earth's gyroscopic frequency has been applied. We have extracted Kimball 2017 from Ref. \cite{kimball2017} and Zhang 2023 from Ref. \cite{xenonNeutronBounds}. Data from Venema 1992 \cite{fortson1992} was extracted by Ref. \cite{hunter2013}. Plots depict constraints on (a) $e$-$n$ and (b) $e$-$p$ couplings for $V_2$, and (c) $e$-$n$ and (d) $e$-$p$ couplings for $V_{11}$. All curves have been calculated using the formalism and coefficients established in Ref. \cite{budker2025}.}
    \label{fig:bounds}
\end{figure*}

Since 2013, two spin-gravity searches were completed which we have re-interpreted using our Earth-spin model. We have analyzed the results of a $^{129}$Xe-$^{131}$Xe spin-gravity experiment completed in 2023 \cite{xenonNeutronBounds}. In the context of our spin model, we use their reported bound, $\beta_z^n$, to extract limits on the $e$-$n$ couplings. Those bounds are reported in Figures \ref{fig:bounds}a and \ref{fig:bounds}c. Ref. \cite{kimball2017} searched for exotic proton spin couplings using the different hyperfine levels of $^{85}$Rb and $^{87}$Rb. This experiment is unique in that its interpretation does not rely on any complex nuclear theory to extract $e$-$p$ bounds. We use their reported orientation-dependent energy bounds, $\beta_z^{p}$, to extract bounds from their results and report them in Figures \ref{fig:bounds}b and \ref{fig:bounds}d. Additionally, we include the bounds extracted from Ref. \cite{fortson1992} by Ref. \cite{hunter2013} in Figure \ref{fig:bounds}.

Figure \ref{fig:bounds} graphs our bounds as functions of $\lambda$ for the $e$-$p$ and $e$-$n$ coupling constants for $V_2$ and $V_{11}$. For these potentials, our results set the most stringent bounds for $\lambda>1$ km. In the long-range/zero-mass limit $\lambda=\infty$, we derive constraints for $V_2$ of $|g_A^eg_A^n| \leq 3.0 \times 10^{-48}$ and $|g_A^eg_A^p| \leq 3.0 \times 10^{-47}$. For $V_{11}$, our massless bounds are $|g_A^eg_V^n|\leq 2.8\times 10^{-20}$ and $|g_A^eg_V^p|\leq 2.7\times 10^{-19}$. The constraints on the axial-axial couplings $g_Ag_A$ are easier to interpret due to the clearly asymptotic shape of the graphs. By contrast, the constraints on the axial-vector couplings $g_Ag_V$ do not represent the tightest constraints for all possible $\lambda$ values. For $\lambda \sim 10^3$ km, we achieve our most stringent bounds of $|g_A^e g_V ^n| \leq 2.2 \times 10^{-25}$ and $|g_A^e g_V ^p| \leq 2.2 \times 10^{-24}$. This suggests that, for $V_{11}$, longer ranges experience strong cancellation from contributions from the far side of Earth. 

\section{Discussion \& Outlook}

Our new constraints on long-range spin-dependent interactions represent the most stringent bounds to-date for the coupling of electron spins to neutron and proton spins. By refining the long-range bound, we further constrain the parameter space in which extensions to the Standard Model may operate. We note that the $e$-$n$ bound on $V_2$ is approximately 17 million times smaller than the gravitational interaction between the electron and neutron. As such, our results may be of interest to some torsion gravity models \cite{kostolecky1998, hammond2002,flambaum2009}.

In this work, we focused on 2 of the 16 possible potentials \cite{dobrescu2006}. We hope in the near future to evaluate bounds on additional potentials. Based on present order-of-magnitude estimates, we anticipate setting new bounds on the $e$-$n$ and $e$-$p$ coupling constants for many of these potentials. We also plan to investigate the possible benefits that might be achieved by incorporating $^{201}$Hg into a future co-magnetometer \cite{fortson1992} and by increasing the spin-relaxation time of Hg \cite{lamoreaux2025}.

%TC:ignore

\section{Acknowledgments}

This work was supported by the National Science Foundation Grant No. PHY-2110523 and was performed in part using the high-performance computing equipment at Amherst College obtained under National Science Foundation Grant No. 2117377. We thank Jim Kubasek for machining support; Brian Crepeau for electronics support; Eric Lindahl, Mikhail Balabas, and Mike Souza for assistance with vapor cell fabrication; and Blayne Heckel and Steve Lamoreaux for helpful conversations. We thank Bek Herz for their contributions to the early phases of this work.

\clearpage

\appendix

\section{Local Systematics}\label{app:localSystematics}

Table \ref{tab:localSystematics} records the 13 local systematic effects that we measure during our data collection process. In the actual data analysis pipeline, we apply our corrections on an octet-by-octet basis. Here, we take a weighted average of individual systematic contributions to our signal using the statistical weights of the octets. This framing helps assess the relative size and variability of the systematics. In the first 7 rows, we consider the effects of laser power and frequency variations expected to arise from AC light effects \cite{happerOpticalPumping}. Rows 8-10 evaluate effects from differences between the pump intensity modulation frequency and the atomic precession frequency (the ``asynchronous pumping effects," or ``APEs"). The ``Quadrupole" correction accounts for a varying quadrupolar magnetic field in our three-cell comagnetometer arrangement. ``Beam Wander" estimates an upper bound on the effect associated with the motion of a laser beam upon table rotation in the presence of a magnetic gradient. ``Horizontal Tip Drift" evaluates how a drift in the $\hat{y}$ component of the magnetic field affects our signal by modifying the noninertial reference frame correction. As this effect is only second-order sensitive for North-South, we have excluded it from the North-South column. The ``Sum" row takes the overall effect of these systematics, which we use to correct the data for each configuration.

\begin{table}[h]
    \centering
    \begin{tabular}{|l|r|r|}\hline
        Systematic & North-South & East-West \\ \hline
        Hg Probe Power & $0.0\pm 0.4$ & $0.0\pm 0.3$ \\
        Hg Pump Power & $-0.1\pm 0.2$ & $0.0\pm 0.2$ \\
        Cs Probe Power & $0.0\pm 0.1$ & $0.0\pm 0.1$ \\
        Cs Pump Power & $-0.9 \pm 4.6$ & $0.2\pm 4.6$ \\
        Cs Probe Frequency & $-0.2\pm1.3$ & $0.2\pm1.4$ \\
        Hg Pump Frequency & $0.0\pm 0.1$ & $0.0\pm0.1$ \\
        Hg Probe Frequency & $0.0\pm 0.0$ & $0.0\pm0.0$ \\
        Hg APEs & $4.6\pm 0.5$ & $0.0\pm 0.4$ \\
        Cs1 APEs & $-1.5\pm 1.7$ & $0.4\pm 1.6$ \\
        Cs2 APEs & $-1.5\pm 1.7$ & $0.4\pm 2.0$ \\
        Quadrupole & $6.5\pm 6.0$ & $-0.7\pm 6.0$ \\
        Beam Wander & $0.0\pm 0.9$ & $0.0\pm 0.9$ \\
        Horizontal Tip Drift & --- &$ -0.8\pm 1.1$ \\  \hline
        Sum & $6.8\pm 8.2$ & $-0.2\pm 8.2$ \\ \hline
    \end{tabular}
    \caption{Local Systematics. Units are Hg nHz.}
    \label{tab:localSystematics}
\end{table}

\section{Global Systematics}\label{app:globalSystematics}

Five ``global" systematic corrections that apply to all of our data are shown in Table \ref{tab:systematicChecks}. Of these five, only the ``Noninertial Reference Frame" significantly alters the central values of our results. Earth is a noninertial reference frame rotating at $f_E=11,605.8$ nHz \cite{nasaAstrodynamicParameters}. The precession frequencies of our atoms shift depending on the projection of our magnetic field along Earth's rotation axis. Assuming perfect alignment of our magnetic field with the horizontal cardinal directions, we expect our North-South difference signals to be modified by
\begin{equation}
    \Delta f_I = 2f_E\left(1-\frac{\gamma_{^{199}\text{Hg}}}{\gamma_\text{Cs,F=4}}\right)\cos(\theta_{L})=17,184.5\text{ nHz}
\end{equation}
where we have used the gyromagnetic ratios $\gamma_{\text{Cs,F=4}}=-3,501.2$ Hz/$\mu$T \cite{csEnergyLevels} and $\gamma_{^{199}\text{Hg}}=7.5901152(13)$Hz/$\mu$T \cite{hgGyro} and our latitude in Amherst $\theta_L=42.3757^\circ$. We expect the difference signal for East-West to be zero. Because of the large size of the North-South offset, it was removed for clarity at the beginning of our analysis. However, our magnetic field is not always perfectly aligned with the cardinal directions. The North-South data is second-order sensitive to horizontal changes, while East-West is first-order sensitive. The global systematics associated with horizontal angular misalignments are quantified in the first row of Table \ref{tab:systematicChecks}. ``Vertical Tip" reflects the uncertainty associated with the alignment of our table rotation axis with $\hat{z}$. ``Vertical Tip Drift" accounts for possible changes in the vertical magnetic field during data collection. The fourth row bounds a possible systematic effect that arises from our main magnetic coil rectifying the 40 MHz radio frequency applied to the double-passed acousto-optic modulator used to pulse the Cs pump beam. Finally, the fifth row is an upper bound on the systematic associated with variations in the optical frequency of the Cs pump laser.

Total systematic error cannot be reliably estimated through forward propagation of uncertainties due to our use of local corrections before a weighted average. Instead, we estimate our total systematic uncertainties by subtracting in quadrature the errors in the corrected and uncorrected data. We retrieve an experiment-wide systematic uncertainty of $2\sigma_{\text{syst}}=18.8$ Hg nHz in East-West and $2\sigma_{\text{syst}}=12.1$ Hg nHz in North-South. The global systematic error exceeds the experiment-wide systematic uncertainty in East-West. This can be accounted for by noting that the local systematic corrections decrease the weighted variance of the mean.

\begin{table}[ht]
    \centering
    \renewcommand{\arraystretch}{1.2}
    \resizebox{\linewidth}{!}{
    \begin{tabular}{|c|r|r|}\hline
        Systematic Error & North-South & East-West \\ \hline
        Noninertial Reference Frame & $17184.5 \pm 0.0$ & $-1.2 \pm 17.7$\\
        Vertical Tip & $0\pm 6.9$ & $0\pm 6.9$ \\
        Vertical Tip Drift & $0 \pm 1.4$ & $0\pm 1.6$ \\
        AOM RF Signal Rectification & $0 \pm 0.6$ & $0\pm 0.5$ \\
        Cs Pump Optical Frequency & $0 \pm 1.9$ & $0 \pm 1.6$ \\ \hline 
        Sum & $17184.5\pm 7.1$ & $-1.2 \pm 19.1$ \\ \hline
    \end{tabular} }
    \caption{Global Systematics. Units are Hg nHz.}
    \label{tab:systematicChecks}
\end{table}

\section{Cross Checks}\label{app:crossChecks}

To investigate possible undiscovered systematics, data were collected in different configurations. We collected data with an applied field of either 1 $\mu$T or 0.5 $\mu$T. We denote these possibilities as ``R" and ``r" respectively. Variation of our result with this change might suggest some non-linearity or precession-frequency sensitivity in our measurement. We also collected data with two opposite orientations of the applied magnetic field with respect to the rotating table. We distinguish these by a ``+" or a ``$-$" attached to ``R" or ``r" when the magnetic field points toward one side of the table or the other. Averaging over data collected in these opposing configurations of the field should largely cancel table-position-dependent effects from our signal. The remaining parameter change was a reversal of the time sequence used for collecting our data. To do this, we reverse the ordering the time sequence of our octets (e.g. from N-S-S-N-S-N-N-S to S-N-N-S-N-S-S-N). We refer to these as ``T+" and ``T$-$" respectively. A change in the data with this reversal might suggest some hysteresis effect in our data acquisition. For each parameter, we took roughly half of our data with each of the two different options. The change in our signal with the reversal of each of these parameters is summarized in Table \ref{tab:configurationComparisons}. Finally, we have also taken data with the cancellation field along North-South turned off, and we found a North-South value of $42 \pm 120$ Hg nHz and an East-West value of $58 \pm 120$ Hg nHz, consistent with our result. Since we regard this data as compromised, we do not include this data in our final dataset. Note that all of the measured deviations are within $2\sigma$ of a null result. We find that the consistency between the data taken in these different configurations lends confidence to our final result. 
 
\begin{table}[h]
    \centering
    \renewcommand{\arraystretch}{1.2}
    \begin{tabular}{|c|r|r|}\hline
        Configuration & North-South & East-West \, \\ \hline
        $+$ vs. $-$ Field & $-78.2\pm89.5$ & $-85.5\pm93.2$ \\
        T$+$ vs. T$-$ & $22.0 \pm 89.9$ & $-43.3 \pm 93.4$ \\
        R vs. r & $-46.0 \pm 91.1$ & $-25.1 \pm 95.1$ \\ \hline
    \end{tabular}
    \caption{Frequency difference between data collected in complementary configurations. All answers are listed in units of Hg nHz.}
    \label{tab:configurationComparisons}
\end{table}

\bibliography{bib}

@article{dobrescu2006,
  author = {Bogdan A. Dobrescu and Irina Mocioiu},
  title = {Spin-Dependent Macroscopic Forces from New Particle Exchange},
  journal = {Journal of High-Energy Physics},
  volume = {2006},
  issue = {11},
  pages = {005},
  year = {2006},
  doi = {10.1088/1126-6708/2006/11/005}
}

@article{budker2025,
  title = {Spin-dependent exotic interactions},
  author = {Cong, Lei and Ji, Wei and Fadeev, Pavel and Ficek, Filip and Jiang, Min and Flambaum, Victor V. and Guan, Haosen and Jackson Kimball, Derek F. and Kozlov, Mikhail G. and Stadnik, Yevgeny V. and Budker, Dmitry},
  journal = {Rev. Mod. Phys.},
  volume = {97},
  issue = {2},
  pages = {025005},
  numpages = {86},
  year = {2025},
  month = {Jun},
  publisher = {American Physical Society},
  doi = {10.1103/RevModPhys.97.025005},
  url = {https://link.aps.org/doi/10.1103/RevModPhys.97.025005}
}

@article{noNetSpin,
  title = {Spherically symmetric Earth models yield no net electron spin},
  author = {Clayburn, N. B. and Glassford, A. and Leiker, A. and Uelmen, T. and Lin, J. F. and Hunter, L. R.},
  journal = {Phys. Rev. D},
  volume = {111},
  issue = {1},
  pages = {015015},
  numpages = {9},
  year = {2025},
  month = {Jan},
  publisher = {American Physical Society},
  doi = {10.1103/PhysRevD.111.015015},
  url = {https://link.aps.org/doi/10.1103/PhysRevD.111.015015}
}

@article{waxCoating,
  title = {Polarized Alkali-Metal Vapor with Minute-Long Transverse Spin-Relaxation Time},
  author = {Balabas, M. V. and Karaulanov, T. and Ledbetter, M. P. and Budker, D.},
  journal = {Phys. Rev. Lett.},
  volume = {105},
  issue = {7},
  pages = {070801},
  numpages = {4},
  year = {2010},
  month = {Aug},
  publisher = {American Physical Society},
  doi = {10.1103/PhysRevLett.105.070801},
  url = {https://link.aps.org/doi/10.1103/PhysRevLett.105.070801}
}

@article{lamoreaux2025,
    author = {Steven K. Lamoreaux},
    title = {Surface and chemical effects on $^{199}$Hg spin polarization relaxation in optically pumped magnetometers},
    journal = {Academia Quantum},
    volume = {2},
    issue = {3},
    pages = {7840},
    year = {2025},
    doi = {10.20935/AcadQuant7840},
    url = {https://www.academia.edu/3064-979X/2/3/10.20935/AcadQuant7840}
}

@article{ang2014,
  title = {Using Geoelectrons to Search for Velocity-Dependent Spin-Spin Interactions},
  author = {Hunter, L. R. and Ang, D. G.},
  journal = {Phys. Rev. Lett.},
  volume = {112},
  issue = {9},
  pages = {091803},
  numpages = {5},
  year = {2014},
  month = {Mar},
  publisher = {American Physical Society},
  doi = {10.1103/PhysRevLett.112.091803},
  url = {https://link.aps.org/doi/10.1103/PhysRevLett.112.091803}
}

@article{hunter2013,
  author = {Larry Hunter and Joel Gordon and Stephen Peck and Daniel Ang and Jung-Fu Lin},
  title = {Using the Earth as a Polarized Electron Source to Search for Long-Range Spin-Spin Interactions},
  journal = {Science},
  volume = {339},
  number = {6122},
  pages = {928-932},
  year = {2013},
  doi = {10.1126/science.1227460},
  url = {https://www.science.org/doi/abs/10.1126/science.1227460}
}

@article{hgGyro,
  title = {A measurement of the neutron to 199Hg magnetic moment ratio},
  journal = {Physics Letters B},
  volume = {739},
  pages = {128-132},
  year = {2014},
  issn = {0370-2693},
  doi = {https://doi.org/10.1016/j.physletb.2014.10.046},
  url = {https://www.sciencedirect.com/science/article/pii/S0370269314007692},
  author = {S. Afach and C.A. Baker and G. Ban and G. Bison and K. Bodek and M. Burghoff and Z. Chowdhuri and M. Daum and M. Fertl and B. Franke and others},
}

@article{seattleExperiment,
  author = {Yang, Yucheng and Wu, Teng and Chen, Jingbiao and Peng, Xiang and Guo, Hong},
  date = {2021/02/25},
  doi = {10.1007/s00340-021-07594-w},
  id = {Yang2021},
  isbn = {1432-0649},
  journal = {Applied Physics B},
  number = {3},
  pages = {40},
  title = {All-optical single-species cesium atomic comagnetometer with optical free induction decay detection},
  url = {https://doi.org/10.1007/s00340-021-07594-w},
  volume = {127},
  year = {2021}
}

@article{hgNuclearTheory,
  author = {D.F. Jackson Kimball},
  title = {Nuclear spin content and constraints on exotic spin-dependent couplings},
  journal = {New Journal of Physics},
  year = {2015}
}

@techreport{csEnergyLevels,
  author = {Daniel Adam Steck},
  title = {Cesium D Line Data},
  institution = {Oregon Center for Optics and Department of Physics, Univeristy of Oregon},
  year = {1998},
  url = {https://steck.us/alkalidata/cesiumnumbers.pdf}
}

@article{romalisHg,
  title = {Techniques used to search for a permanent electric dipole moment of the ${}^{199}$Hg atom and the implications for $\mathit{CP}$ violation},
  author = {Swallows, M. D. and Loftus, T. H. and Griffith, W. C. and Heckel, B. R. and Fortson, E. N. and Romalis, M. V.},
  journal = {Phys. Rev. A},
  volume = {87},
  issue = {1},
  pages = {012102},
  numpages = {24},
  year = {2013},
  month = {Jan},
  publisher = {American Physical Society},
  doi = {10.1103/PhysRevA.87.012102},
  url = {https://link.aps.org/doi/10.1103/PhysRevA.87.012102}
}

@article{happerOpticalPumping,
  title = {Optical Pumping},
  author = {Happer, William},
  journal = {Rev. Mod. Phys.},
  volume = {44},
  issue = {2},
  pages = {169--249},
  numpages = {0},
  year = {1972},
  month = {Apr},
  publisher = {American Physical Society},
  doi = {10.1103/RevModPhys.44.169},
  url = {https://link.aps.org/doi/10.1103/RevModPhys.44.169}
}

@article{xenonNeutronBounds,
  title = {Search for Spin-Dependent Gravitational Interactions at Earth Range},
  author = {Zhang, S.-B. and Ba, Z.-L. and Ning, D.-H. and Zhai, N.-F. and Lu, Z.-T. and Sheng, D.},
  journal = {Phys. Rev. Lett.},
  volume = {130},
  issue = {20},
  pages = {201401},
  numpages = {6},
  year = {2023},
  month = {May},
  publisher = {American Physical Society},
  doi = {10.1103/PhysRevLett.130.201401},
  url = {https://link.aps.org/doi/10.1103/PhysRevLett.130.201401}
}

@article{fortson1992,
  title = {Search for a coupling of the Earth's gravitational field to nuclear spins in atomic mercury},
  author = {Venema, B. J. and Majumder, P. K. and Lamoreaux, S. K. and Heckel, B. R. and Fortson, E. N.},
  journal = {Phys. Rev. Lett.},
  volume = {68},
  issue = {2},
  pages = {135--138},
  numpages = {0},
  year = {1992},
  month = {Jan},
  publisher = {American Physical Society},
  doi = {10.1103/PhysRevLett.68.135},
  url = {https://link.aps.org/doi/10.1103/PhysRevLett.68.135}
}

@article{heckel2008,
  title = {Preferred-frame and $CP$-violation tests with polarized electrons},
  author = {Heckel, B. R. and Adelberger, E. G. and Cramer, C. E. and Cook, T. S. and Schlamminger, S. and Schmidt, U.},
  journal = {Phys. Rev. D},
  volume = {78},
  issue = {9},
  pages = {092006},
  numpages = {23},
  year = {2008},
  month = {Nov},
  publisher = {American Physical Society},
  doi = {10.1103/PhysRevD.78.092006},
  url = {https://link.aps.org/doi/10.1103/PhysRevD.78.092006}
}

@article{darkPhotonTheory,
  title = {Direct detection constraints on dark photon dark matter},
  author = {Haipeng An and Maxim Pospelov and Josef Pradler and Adam Ritz},
  journal = {Physics Letters B},
  volume = {747},
  pages = {331-338},
  year = {2015},
  issn = {0370-2693},
  doi = {https://doi.org/10.1016/j.physletb.2015.06.018},
  url = {https://www.sciencedirect.com/science/article/pii/S0370269315004402}
}

@article{axionOriginalPaper,
  title = {$\mathrm{CP}$ Conservation in the Presence of Pseudoparticles},
  author = {Peccei, R. D. and Quinn, Helen R.},
  journal = {Phys. Rev. Lett.},
  volume = {38},
  issue = {25},
  pages = {1440--1443},
  numpages = {0},
  year = {1977},
  month = {Jun},
  publisher = {American Physical Society},
  doi = {10.1103/PhysRevLett.38.1440},
  url = {https://link.aps.org/doi/10.1103/PhysRevLett.38.1440}
}

@article{alpsPaper,
   author = "Graham, Peter W. and Irastorza, Igor G. and Lamoreaux, Steven K. and Lindner, Axel and van Bibber, Karl A.",
   title = "Experimental Searches for the Axion and Axion-Like Particles", 
   journal= "Annual Review of Nuclear and Particle Science",
   year = "2015",
   volume = "65",
   number = "Volume 65, 2015",
   pages = "485-514",
   doi = "https://doi.org/10.1146/annurev-nucl-102014-022120",
   url = "https://www.annualreviews.org/content/journals/10.1146/annurev-nucl-102014-022120",
   publisher = "Annual Reviews",
   issn = "1545-4134"
}

@article{paraphotonTheory,
  title = {Massless Gauge Bosons other than the Photon},
  author = {Dobrescu, Bogdan A.},
  journal = {Phys. Rev. Lett.},
  volume = {94},
  issue = {15},
  pages = {151802},
  numpages = {4},
  year = {2005},
  month = {Apr},
  publisher = {American Physical Society},
  doi = {10.1103/PhysRevLett.94.151802},
  url = {https://link.aps.org/doi/10.1103/PhysRevLett.94.151802}
}

@article{clayburn2023,
  title = {Using Earth to search for long-range spin-velocity interactions},
  author = {Clayburn, N. B. and Hunter, L. R.},
  journal = {Phys. Rev. D},
  volume = {108},
  issue = {5},
  pages = {L051701},
  numpages = {6},
  year = {2023},
  month = {Sep},
  publisher = {American Physical Society},
  doi = {10.1103/PhysRevD.108.L051701},
  url = {https://link.aps.org/doi/10.1103/PhysRevD.108.L051701}
}

@article{safronova2018,
  title = {Search for new physics with atoms and molecules},
  author = {Safronova, M. S. and Budker, D. and DeMille, D. and Kimball, Derek F. Jackson and Derevianko, A. and Clark, Charles W.},
  journal = {Rev. Mod. Phys.},
  volume = {90},
  issue = {2},
  pages = {025008},
  numpages = {106},
  year = {2018},
  month = {Jun},
  publisher = {American Physical Society},
  doi = {10.1103/RevModPhys.90.025008},
  url = {https://link.aps.org/doi/10.1103/RevModPhys.90.025008}
}

@article{wu2023,
  title = {New Limits on Exotic Spin-Dependent Interactions at Astronomical Distances},
  author = {Wu, L. Y. and Zhang, K. Y. and Peng, M. and Gong, J. and Yan, H.},
  journal = {Phys. Rev. Lett.},
  volume = {131},
  issue = {9},
  pages = {091002},
  numpages = {7},
  year = {2023},
  month = {Aug},
  publisher = {American Physical Society},
  doi = {10.1103/PhysRevLett.131.091002},
  url = {https://link.aps.org/doi/10.1103/PhysRevLett.131.091002}
}

@article{poddar2023,
  title = {Constraints on monopole-dipole potential from tests of gravity},
  author = {Poddar, Tanmay Kumar and Pachhar, Debashis},
  journal = {Phys. Rev. D},
  volume = {108},
  issue = {10},
  pages = {103024},
  numpages = {13},
  year = {2023},
  month = {Nov},
  publisher = {American Physical Society},
  doi = {10.1103/PhysRevD.108.103024},
  url = {https://link.aps.org/doi/10.1103/PhysRevD.108.103024}
}

@ARTICLE{hammond2002,
  author = {{Hammond}, Richard T.},
  title = {Torsion gravity},
  journal = {Reports on Progress in Physics},
  year = {2002},
  month = {may},
  volume = {65},
  number = {5},
  pages = {599-649},
  doi = {10.1088/0034-4885/65/5/201},
  adsurl = {https://ui.adsabs.harvard.edu/abs/2002RPPh...65..599H}
}

@article{kostolecky1998,
  title = {Lorentz-violating extension of the standard model},
  author = {Colladay, D. and Kosteleck\'y, V. Alan},
  journal = {Phys. Rev. D},
  volume = {58},
  issue = {11},
  pages = {116002},
  numpages = {23},
  year = {1998},
  month = {Oct},
  publisher = {American Physical Society},
  doi = {10.1103/PhysRevD.58.116002},
  url = {https://link.aps.org/doi/10.1103/PhysRevD.58.116002}
}

@article{flambaum2009,
  title = {Scalar-tensor theories with pseudoscalar couplings},
  author = {Flambaum, Victor and Lambert, Simon and Pospelov, Maxim},
  journal = {Phys. Rev. D},
  volume = {80},
  issue = {10},
  pages = {105021},
  numpages = {8},
  year = {2009},
  month = {Nov},
  publisher = {American Physical Society},
  doi = {10.1103/PhysRevD.80.105021},
  url = {https://link.aps.org/doi/10.1103/PhysRevD.80.105021}
}

@article{huang2025,
  title = {Hunting for exotic bosons with flying quantum sensors in space},
  author = {Huang, Xingming and Wang, Yuanhong and Kang, Xiang and Li, Jiaxi and Su, Haowen and Wang, Zehao and Lin, Qing and Zheng, Wenqiang and Sun, Yuan and Liu, Liang and Jiang, Min and Peng, Xinhua and Zhao, Zhengguo and Du, Jiangfeng},
  journal = {Phys. Rev. D},
  volume = {112},
  issue = {9},
  pages = {095015},
  numpages = {23},
  year = {2025},
  month = {Nov},
  publisher = {American Physical Society},
  doi = {10.1103/39cs-rn8k},
  url = {https://link.aps.org/doi/10.1103/39cs-rn8k}
}

@Article{lai2026,
  title = {Potential of Constraining the Fifth Force Using the Earth as a Spin and Mass Source from Space},
  journal = {Chin. Phys. Lett.},
  volume = {43},
  number = {3},
  pages = {030202-030202},
  year = {2026},
  doi = {10.1088/0256-307X/43/3/030202},	
  url = {http://cpl.iphy.ac.cn/en/article/doi/10.1088/0256-307X/43/3/030202},
  author = {Zheng-Ting Lai and Jun-Xu Lu and Li-Sheng Geng and Kai Wei and Wei Ji}
}

@article{kimball2017,
  title = {Constraints on long-range spin-gravity and monopole-dipole couplings of the proton},
  author = {Jackson Kimball, Derek F. and Dudley, Jordan and Li, Yan and Patel, Dilan and Valdez, Julian},
  journal = {Phys. Rev. D},
  volume = {96},
  issue = {7},
  pages = {075004},
  numpages = {10},
  year = {2017},
  month = {Oct},
  publisher = {American Physical Society},
  doi = {10.1103/PhysRevD.96.075004},
  url = {https://link.aps.org/doi/10.1103/PhysRevD.96.075004}
}

@article{peck2016,
  title = {Using tensor light shifts to measure and cancel a cell's quadrupolar frequency shift},
  author = {Peck, S. K. and Lane, N. and Ang, D. G. and Hunter, L. R.},
  journal = {Phys. Rev. A},
  volume = {93},
  issue = {2},
  pages = {023426},
  numpages = {8},
  year = {2016},
  month = {Feb},
  publisher = {American Physical Society},
  doi = {10.1103/PhysRevA.93.023426},
  url = {https://link.aps.org/doi/10.1103/PhysRevA.93.023426}
}

@techreport{nasaAstrodynamicParameters,
  author = {NASA},
  title = {Astrodynamic Parameters},
  institution = {Jet Propoulsion Laboratory, California Institute of Technology},
  year = {2026},
  url = {https://ssd.jpl.nasa.gov/astro_par.html}
}

@book{weightedavg,
  title = {Data Reduction and Error Analysis for the Physical Sciences},
  edition = {Third},
  author = {Bevington, P. R. and Robinson, D. K.},
  publisher = {McGraw-Hill},
  year = {2003},
  pages = {58}
}

%TC:endignore

\end{document}